\documentstyle{elsartwb}

\input{epsf}

\begin{document}

\begin{frontmatter}

\hfill INP 1836/PH

\title{Chiral restoration in effective
quark models with non-local interactions\thanksref{grant}}
\thanks[grant]{Supported by the State Committee for Scientific
Research, grant 2P03B-080-12.}
\author{{Barbara Szczerbi\'{n}ska\thanksref{USC}${}^1$}
and {Wojciech Broniowski\thanksref{email}${}^2$}}
\thanks[USC]{Present address: Graduate School,
Department of Physics and Astronomy,
University of South Carolina, Columbia, SC 29108, U.S.A.}
\thanks[email]{broniows@solaris.ifj.edu.pl}
\address{H. Niewodnicza\'{n}ski Institute of Nuclear Physics,
         PL-31342 Krak\'{o}w, POLAND}

\begin{abstract}
Chiral restoration at finite temperatures is studied in chiral quark models
with non-local regulators. At the leading-$N_c$ level we find transition
temperature of the order 100MeV. Meson-loop contributions are also analyzed
and found to have a very small effect.
\end{abstract}

\end{frontmatter}

Recently, considerable activity has been focused on chiral quark models with 
{\em non-local} interactions \cite
{Ripka97,Cahill87,Holdom89,Ball90,Krewald92,%
Ripka93,BowlerB,PlantB,mitia:rev,Bijnens,Blaschke,BlaRob,sch1,sch2,%
solnonl,bled99,coim99wb}%
. In this approach interactions depend on the momenta carried by the quarks,
which leads to a momentum-dependent quark mass, generated by spontaneous
breaking of the chiral symmetry. There are several important reasons why it
is worthwhile to consider such models: (1) Non-locality arises naturally in
the most sound and successful approaches to low-energy quark dynamics,
namely the instanton-liquid model \cite{Diakonov86,instant1:rev,instant2:rev}
and the Schwinger-Dyson resummation techniques \cite{Roberts92}. Hence, we
should deal with non-local regulators from the outset and models with local
interactions (the Nambu--Jona-Lasinio model and its progeny) may be viewed
as simplifications to the non-local models. (2) The non-local regularization
leads to preservation of anomalies \cite{Ripka93,Arrio} and to correct
charge quantization \cite{solnonl,bled99}. With local methods, such as the
proper-time regularization or the quark-loop momentum cut-off \cite
{mitia:rev,Bijnens,Goeke96,Reinhardt96} the matching of anomalies can only
be achieved if the finite (anomalous) part of the Euclidean action is left
unregularized, and the infinite (non-anomalous) part is regularized. When
both parts are regularized, anomalies are violated substantially \cite
{krs,bbs} and the theory is not consistent. The different treatment of the
anomalous and non-anomalous parts of the action is rather artificial and it
is quite appealing that with non-local regulators both parts of the action
can be treated in a unified way. (3) With non-local interactions the
effective action is finite to all orders in the loop expansion ({\em i.e.}
the $1/N_{c}$ expansion). In particular, the meson loops are finite and it
is not necessary to introduce additional cut-offs, as was the case of local
models \cite{Tegen95,loops:NJL,Temp}. As a result, the non-local models have
more predictive power. (4) In the baryonic sector the non-local models are
capable of producing stable solitons \cite{solnonl} without the extra
constraint that forces the $\sigma $ and $\pi $ fields to lie on the chiral
circle. Such a constraint is external to the known derivations of effective
chiral quark models.

In view of the above-listed advantages of the non-local models over the
conventional Nambu--Jona-Lasinio-like models it is desired to check other
applications of non-local effective chiral theories. In this paper we
investigate the restoration of chiral symmetry at finite temperatures and
vanishing baryon density. The basic quantity of our study is the quark
condensate, which in the chiral limit of the vanishing current quark mass is
the order parameter of chiral symmetry. We show that chiral restoration is a
second-order phase transition, as in the local models, and it occurs at
temperatures $T_{c}\sim 100{\rm MeV}$, which is somewhat less than in local
theories, where the typical $T_{c}$ is around $150{\rm MeV}$. We also
investigate the dependence of the critical temperature $T_{c}$ on the
detailed shape of the non-local regulator and find it to be weak. With a
finite current quark mass the chiral transition is, as usual, a smooth
cross-over. In addition, we show that the effects of meson-loops
(higher-order effects in the $1/N_{c}$ expansion) of the quark condensate
are very small in non-local models. 

We begin with a brief review of the model. Our starting point is the
two-flavor effective non-local Euclidean action in the bosonized form 
\begin{equation}
\begin{array}{c}
I=-{\rm Tr}\ln \left( -i\gamma _{\mu }\partial ^{\mu }+m+r\Phi \Gamma _{\Phi
}r\right) +\frac{a^{2}}{2}\int d^{4}x(\Phi _{0}^{2}+{\bf \Phi ^{2}}),
\end{array}
\label{Ieff}
\end{equation}
where $\frac{1}{a^{2}}$ has the meaning of the coupling constant of the
four-quark interactions, $m$ is the current quark mass, $\Phi =\left( \Phi
_{0},{\bf \Phi }_{a}\right) $ are the mean meson fields with quantum numbers
of the $\sigma $ meson and the pions, respectively. The fields $\Phi $ are
local in coordinate space. Matrices $\Gamma _{\Phi }=\left( \Gamma _{\Phi
_{0}},\Gamma _{\Phi _{a}}\right) $ are acting in spin and flavor spaces,
with $\Gamma _{\Phi _{0}}=1$ and $\Gamma _{{\bf \Phi }_{a}}=i\gamma _{5}{\bf %
\tau }^{a}$. The regulator operator $r$ is local in momentum space, {\em i.e.%
} $\langle p|r|p^{\prime }\rangle =\delta (p-p^{\prime })r(p)$. It
introduces a cut-off scale $\Lambda $, limiting appropriately momenta of the
quarks. The symbol ${\rm Tr}$ denotes the full trace, {\em i.e.} the
functional trace and the matrix traces over the Dirac, flavor and color
degrees of freedom.

Through the use of the Euler-Lagrange equations for the $\Phi $ field it is
straightforward to show that the action (\ref{Ieff}) is equivalent to the
model where the four-quark interaction has a separable form \cite
{BowlerB,PlantB}, namely 
\begin{eqnarray}
L_{{\rm int}} &=&-\frac{1}{2a^{2}}\int \frac{d^{4}p_{1}}{(2\pi )^{4}}\int 
\frac{d^{4}p_{2}}{(2\pi )^{4}}\int \frac{d^{4}p_{3}}{(2\pi )^{4}}\int \frac{%
d^{4}p_{4}}{(2\pi )^{4}}\times   \label{lint} \\
&&\delta ^{4}(p_{1}+p_{2}+p_{3}+p_{4})\overline{\psi }(p_{1})r(p_{1})\Gamma
_{\Phi }r(p_{2})\psi (p_{2})\overline{\psi }(p_{3})r(p_{3})\Gamma _{\Phi
}r(p_{4})\psi (p_{4}).  \nonumber
\end{eqnarray}
This form has a very clear interpretation via Feynman diagrams: the quark
vertices carry a regulator on each quark line entering the vertex.

In the instanton-liquid model the regulator can be evaluated for low
Euclidean momenta. It has the form \cite
{Diakonov86,instant1:rev,instant2:rev} 
\begin{equation}
r\left( p^{2}\right) =-\frac{z}{2}\frac{d}{dz}\left( I_{0}\left( z\right)
K_{0}\left( z\right) -I_{1}\left( z\right) K_{1}\left( z\right) \right)
,\quad z=\frac{\sqrt{p^{2}}\rho }{2},  \label{Rinst}
\end{equation}
where $I$ and $K$ are the modified Bessel functions, and $\rho $ is the
average instanton size, of the order of (600MeV)$^{-1}$. We introduce $%
\Lambda =2/\rho $, which is therefore of the order of $1.2{\rm GeV}$. We
will also study an approximate formula for the regulator \cite
{BowlerB,PlantB}, given by the Gaussian function:
\begin{equation}
r\left( p^{2}\right) =\exp \left( -\frac{p^{2}}{2\Lambda ^{2}}\right)
\end{equation}

We first analyze the model at the one-quark-loop level. In the vacuum the
values of the fields $\Phi _{0}$ and ${\bf \Phi }_{a}$ are obtained from the
stationary-point conditions: $\frac{\delta I}{\delta {\bf \Phi }^{a}}=0$ and
$\frac{\delta I}{\delta \Phi _{0}}=0$. The first condition is trivially
fulfilled for ${\bf \Phi }_{a}=0$, and the second one has a nontrivial
solution $\Phi _{0}=S_{0}$, which satisfies the equation
\begin{equation}
a^{2}=2\nu \left( g\left( S_{0}\right) +\frac{m}{S_{0}}g^{\prime }\left(
S_{0}\right) \right) ,  \label{varsigq}
\end{equation}
with $g\left( S_{0}\right) $ and $g^{\prime }\left( S_{0}\right) $ given by
the following integrals: 
\begin{equation}
g\left( S_{0}\right) =\int \frac{d^{4}k}{(2\pi )^{4}}\frac{R^{2}(k^{2})}{%
D(k^{2})}  \label{gfun}
\end{equation}
\begin{equation}
g^{\prime }\left( S_{0}\right) =\int \frac{d^{4}k}{(2\pi )^{4}}\frac{R(k^{2})%
}{D(k^{2})}  \label{gpfun}
\end{equation}
Above, we have used the notation 
\begin{eqnarray}
&&\nu =2N_{c}N_{f},\quad D(k^{2})=k^{2}+{\cal M}^{2}(k^{2}),  \nonumber \\
&&{\cal M}(k^{2})=R(k^{2})S_{0}+m,\quad R(k^{2})=r^{2}(k^{2}),
\end{eqnarray}
where $N_{c}=3$ is the number of colors and $N_{f}=2$ is the number of
flavors. The quark condensate (for a single flavor) is defined as $%
\left\langle \overline{q}q\right\rangle \equiv {\frac{1}{N_{f}}}\frac{\delta
I}{\delta m}$ , which explicitly gives 
\begin{equation}
\left\langle \overline{q}q\right\rangle =-4N_{c}\int \frac{d^{4}k}{(2\pi
)^{4}}\frac{{\cal M}(k^{2})}{D(k^{2})}.  \label{condM}
\end{equation}
If $m\neq 0$, the above expression is ultraviolet divergent. However, in
this case one should subtract the perturbative vacuum part from (\ref{condM}%
),{\em \ i.e.} the piece with $S_{0}=0$. The results is well-defined: 
\begin{eqnarray}
\left\langle \overline{q}q\right\rangle  &\longrightarrow &\left\langle 
\overline{q}q\right\rangle -\left\langle \overline{q}q\right\rangle
_{S_{0}=0}=-4N_{c}\int \frac{d^{4}k}{(2\pi )^{4}}\left( \frac{{\cal M}(k^{2})%
}{k^{2}+{\cal M}^{2}(k^{2})}-\frac{m}{k^{2}+m^{2}}\right) =  \nonumber \\
&=&-4N_{c}S_{0}g^{\prime }\left( S_{0}\right) +4N_{c}m\int \frac{d^{4}k}{%
(2\pi )^{4}}\frac{\left( 2m+R(k^{2})S_{0}\right) R(k^{2})S_{0}}{%
D(k^{2})\left( k^{2}+m^{2}\right) }=  \nonumber \\
&=&-4N_{c}S_{0}g^{\prime }\left( S_{0}\right) +{\cal O}\left( m\right) 
\label{conden}
\end{eqnarray}
The model has three parameters: the cut-off $\Lambda $, the current quark
mass $m$ and the coupling constant $\frac{1}{a^{2}}$. Two conditions for
parameters are obtained by demanding that the pion decay constant, $F_{\pi }$%
, and the pion mass, $m_{\pi }$, assume their physical values: $F_{\pi }=93%
{\rm MeV}$, and $m_{\pi }=139{\rm MeV}$. To leading order in $m$ the
expressions for these quantities in our model are \cite{BowlerB,PlantB}:
\begin{equation}
{F_{\pi }}^{2}=\nu S_{0}^{2}\int \frac{d^{4}k}{(2\pi )^{4}}\frac{%
R^{2}(k^{2})-k^{2}R(k^{2})\frac{dR(k^{2})}{d(k^{2})}+k^{4}\left( \frac{%
dR(k^{2})}{d(k^{2})}\right) ^{2}}{D^{2}(k^{2})}\;,  \label{Fpi}
\end{equation}
\begin{equation}
m_{\pi }^{2}=\frac{2\nu S_{0}mg^{\prime }(S_{0})}{F_{\pi }^{2}}.
\label{mpig}
\end{equation}
We note that the elimination of $g^{\prime }(S_{0})$ from r.h.s of the
expression (\ref{mpig}), with help of the equation (\ref{conden}), leads to 
\begin{equation}
m_{\pi }^{2}=-\frac{m\left\langle \overline{q}q\right\rangle }{F_{\pi }^{2}},
\label{mpiqq}
\end{equation}
in which we recognize as the Gell-Mann-Oakes-Renner relation. One parameter
of the model is left free. Through the use of the stationary-point condition
we can trade it for the value of $S_{0}$, which now is treated as the only
remaining parameter. In principle we could still fit it, making use another
helpful quantity, namely the gluon condensate. In our model it can be
written as \cite{Weiss}
\begin{equation}
\left\langle \frac{\alpha _{s}G^{\mu \nu }G_{\mu \nu }}{\pi }\right\rangle =%
\frac{8}{N_{f}}a^{2}S_{0}^{2}-m\langle \overline{q}q\rangle .  \label{glucon}
\end{equation}
However, the dependence of the gluon condensate on $S_{0}$ (with $F_{\pi }$
and $m_{\pi }$ fixed at physical values) is very weak, which makes the fit
impossible in practice due to a large uncertainty in our knowledge of $%
\left\langle \alpha _{s}/\pi \,G^{\mu \nu }G_{\mu \nu }\right\rangle $.
Reasonable parameters yield the value of the quark condensate $\langle 
\overline{q}q\rangle \sim (250{\rm MeV})^{3}$, the current quark mass $m\sim
5{\rm MeV}$ and the gluon condensate $\left\langle \alpha _{s}/\pi \,G^{\mu
\nu }G_{\mu \nu }\right\rangle \sim (330\pm 30{\rm MeV})^{4}$. This favors $%
S_{0}/\Lambda $ in the range $0.2-0.5$.

Now we pass to the analysis of the temperature dependence of the quark
condensate. For this purpose we use the Matsubara formalism, which leads to
the following simple thermalization rule for one-quark-loop integrals 
\begin{equation}
\int \frac{d^{4}k}{\left( 2\pi \right) ^{4}}f(k)=\int \frac{dk_{0}}{2\pi }%
\int \frac{d^{3}k}{\left( 2\pi \right) ^{3}}f(k_{0},\vec{k})\longrightarrow
T\sum_{j=-\infty }^{\infty }\int \frac{d^{3}k}{\left( 2\pi \right) ^{3}}%
f(E_{j},\vec{k}),  \label{temp}
\end{equation}
where $T$ is the temperature, $f$ is the integrand of the quark loop, and $k$
denotes the momentum running around the loop. The summation runs over the
appropriate fermionic Matsubara frequencies, $E_{j}=(2j+1)\pi T$. After
thermalization and integration of the angular dependence, the equations (\ref
{gfun}) and (\ref{gpfun} ) acquire the following finite-temperature form: 
\begin{equation}
g_{T}\left( S_{0}\right) =\frac{1}{\pi ^{2}}T\sum_{n=0}^{\infty }\int dk%
\frac{k^{2}R^{2}(k^{2}+E_{n}^{2})}{k^{2}+E_{n}^{2}+{\cal M}%
^{2}(k^{2}+E_{n}^{2})},  \label{gfunT}
\end{equation}
\begin{equation}
g_{T}^{\prime }\left( S_{0}\right) =\frac{1}{\pi ^{2}}T\sum_{n=0}^{\infty
}\int dk\frac{k^{2}R(k^{2}+E_{n}^{2})}{k^{2}+E_{n}^{2}+{\cal M}%
^{2}(k^{2}+E_{n}^{2})}.  \label{gpfunT}
\end{equation}
Of course the value of $S_{0}$ depends on $T$ via the finite-temperature
stationary-point condition: 
\begin{equation}
a^{2}=2\nu \left( g_{T}\left( S_{0}\right) +\frac{m}{S_{0}}g_{T}^{\prime
}\left( S_{0}\right) \right) ,
\end{equation}
where now $k$ denotes the length of the 3-momentum $\vec{k}$. The
finite-temperature expression for the quark condensate is 
\begin{eqnarray}
\langle \overline{q}q\rangle _{T} &=&-4N_{c}S_{0}g_{T}^{\prime }\left(
S_{0}\right) + \\
&&\frac{4N_{c}m}{\pi ^{2}}T\sum_{n=0}^{\infty }\int dk\frac{k^{2}\left(
2m+R(k^{2}+E_{n}^{2})S_{0}\right) R(k^{2}+E_{n}^{2})S_{0}}{\left(
k^{2}+E_{n}^{2}+{\cal M}(k^{2}+E_{n}^{2})^{2}\right) \left(
k^{2}+E_{n}^{2}+m^{2}\right) },  \nonumber  \label{condT}
\end{eqnarray}

\begin{table}[t]
\caption{Results for the two considered regulators and various parameter
sets. The first column is the ratio $S_{0}/\Lambda $, the next three columns
display the values of the model parameters $\Lambda $, $a$, and $m$,
adjusted in such a way that $F_{\pi }=93{\rm MeV}$ and $m_{\pi }=139{\rm MeV}
$ for the given $S_{0}/\Lambda $. The next two columns show the calculated
values of $S_{0}$ and $\langle -\overline{q}q\rangle ^{\frac{1}{3}}$ at $T=0$%
. The last three columns give the results obtained in the strict chiral
limit, with $m$ set to $0$, and $\Lambda $ and $a$ from columns (2--3). The
quantities $S_{0,m=0}$ and $\langle -\overline{q}q\rangle _{m=0}^{\frac{1}{3}%
}$ denote the values of the scalar field and the quark condensate for this
case and for $T=0$, and $T_{{\rm crit}}$ denotes the critical temperature of
the second-order chiral phase transition. }
\begin{centering}
\begin{tabular}{|c|c|c|c||c|c||c|c|c|}
\hline
$\frac {S_0}\Lambda $ & $\Lambda $ & $a$ & $m$ & $S_0$ & $\langle -\overline{%
q}q\rangle ^{\frac 13}$ & $S_{0,m=0}$ & $\langle -\overline{q}q\rangle
_{m=0}^{\frac 13}$ & $T_{{\rm crit}}$ \\ 
& MeV & MeV & MeV & MeV & MeV & MeV & MeV & MeV \\ \hline
\multicolumn{9}{|l|}{Instanton-model regulator} \\ \hline
0.2 & 1506 & 219 & 3.5 & 301 & 287 & 256 & 274 & 102 \\
0.3 & 1191 & 165 & 4.9 & 357 & 256 & 310 & 246 & 101 \\ 
0.4 & 1019 & 134 & 6.2 & 408 & 238 & 358 & 229 & 100 \\ 
0.5 & 907 & 115 & 7.3 & 454 & 225 & 401 & 218 & 99 \\ \hline
\multicolumn{9}{|l|}{Gaussian regulator} \\ \hline
0.2 & 1155 & 296 & 3.9 & 231 & 290 & 192 & 278 & 106 \\ 
0.3 & 894 & 218 & 5.9 & 268 & 250 & 230 & 243 & 105 \\ 
0.4 & 754 & 176 & 7.7 & 302 & 227 & 263 & 221 & 104 \\ 
0.5 & 666 & 149 & 9.4 & 333 & 211 & 295 & 207 & 103 \\ \hline
\end{tabular}
\end{centering}
\end{table}

There is an important remark to be made in this place. The described way of
thermalization, typically done in effective chiral quark models, tacitly
assumes that the model parameters (here $a$ and $\Lambda $) do not
explicitly depend on $T$. There is no sound reason for this assumption in
our effective theory, except for simplicity. Allowing
temperature dependence of model parameters influences the values of $T_{c}$.
We will return to this question shortly.

Table 1 shows our numerical results, with the meaning of various quantities
given in the caption. The striking feature here is the very weak dependence
of the critical temperature $T_{c}$ on the model parameters and on the
regulator used. We find $T_{c}\sim 100{\rm MeV}$, which is somewhat less
than in conventional Nambu--Jona-Lasinio model, where it is of the order of $%
120-150{\rm MeV}$. The reason for this lower value is the following. The
integrand of the expression for $g_{T}$ in local models is simply $%
k^{2}/(k^{2}+E_{n}^{2}+M^{2})$, where $M$ is the quark mass, instead of our
expression $%
k^{2}R^{2}(k^{2}+E_{n}^{2})/(k^{2}+E_{n}^{2}+M^{2}R^{2}(k^{2}+E_{n}^{2}))$.
The presence of the regulator function $R$ causes the much more rapid
decrease of the integrand with the temperature $T$, which enters through the
Matsubara energies $E_{n}$. The more rapid decrease of $g_{T}$ with $T$
leads directly to a lower critical temperature $T_{c}$ as compared to the
local models. 
\begin{figure}[tbp]
\vspace{-1cm} \epsfxsize = 9cm \centerline{\epsfbox{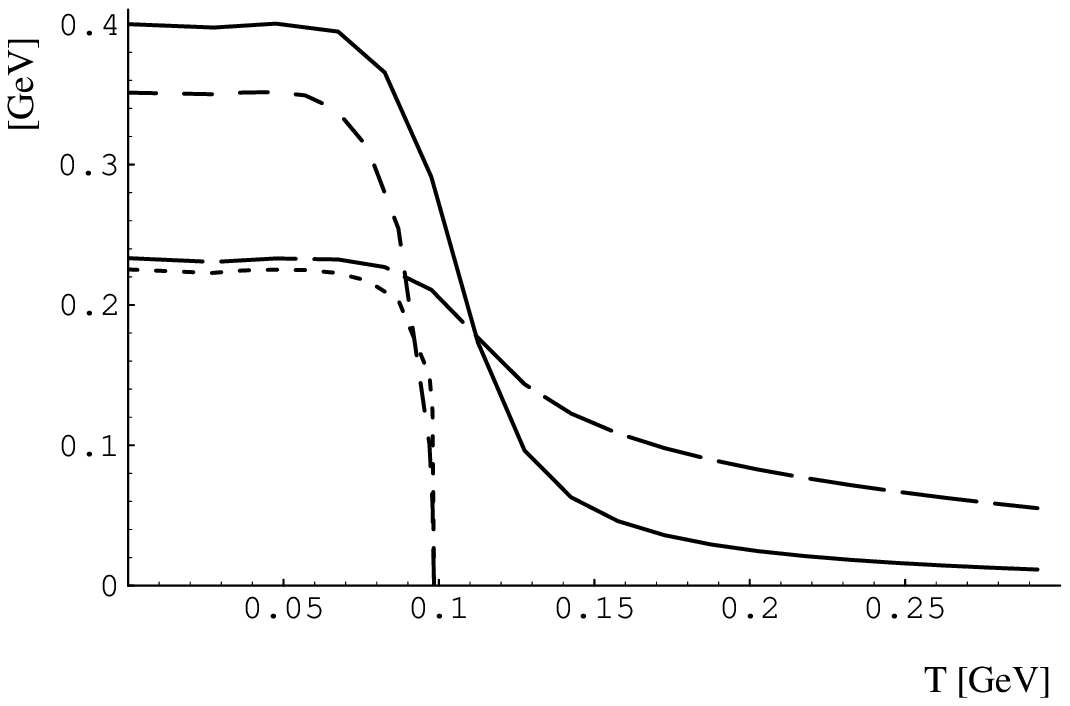}} \vspace{-1cm}
\caption{The temperature dependence of $S_{0}$ and $\langle -\overline{q}%
q\rangle ^{\frac{1}{3}}$ for the instanton-liquid-model regulator with $a=134%
{\rm MeV}$, $\Lambda =1019{\rm MeV}$ and $m=6.19{\rm MeV}$ or $m=0$. The
meaning if the curves is as follows: $S_{0}$ -- solid line, $\langle -%
\overline{q}q\rangle ^{\frac{1}{3}}$ -- long-dash line, $S_{0,m=0}$ --
medium-dash line, $\langle -\overline{q}q\rangle _{m=0}^{\frac{1}{3}}$ --
short-dash line. Note considerable difference between the strict chiral
limit and the finite-$m$ case.}
\label{fig:RYS1}
\end{figure}
One should bare in mind here that any explicit dependence of $a$ on $T$, not
considered here, would influence the values of $T_{c}$. For instance, an
increase of $a$ by 20\% (equivalent to decreasing the coupling constant of
the four-quark interaction) leads to an increase of $T_{c}$ to values around 
$120{\rm MeV}$. We note that calculations in instanton-based models of
the QCD vacuum
\cite{instant1:rev,Diamir,Nowak:inst,Zahed:inst} yield
values of $T_c$ in the range 125--175MeV.

To summarize this part, we note that at the one-quark-loop level the
behavior of the non-local model is qualitatively not different from the
local models. We find the chiral phase transition, which is second-order in
the strict chiral limit, and the corresponding critical temperature is very
weakly dependent on the details of the model.

In the remaining part of this paper we consider meson loops effects. Such
effects, although suppressed by $1/N_{c}$, are expected to be important due
to the low mass of the pion. We follow the method described in Refs. \cite
{Tegen95,loops:NJL}, where the meson loops have been introduced in effective
chiral models in a consistent ({\em i.e.} symmetry-conserving) way. As
mentioned in the introduction, conventional Nambu--Jona-Lasinio models
require an extra regulator for the meson loop -- making the quark loop
finite via momentum cut-off, the proper-time or Pauli-Villars method, still
leaves the momentum in the meson loop unconstrained. With non-local
regulators this is no longer case. The interactions (\ref{lint}) cut-off any
momentum flowing through a quark vertex. As a result, calculation of a
diagram with any number of loops is finite.

\begin{figure}[b]
\epsfxsize = 9cm \centerline{\epsfbox{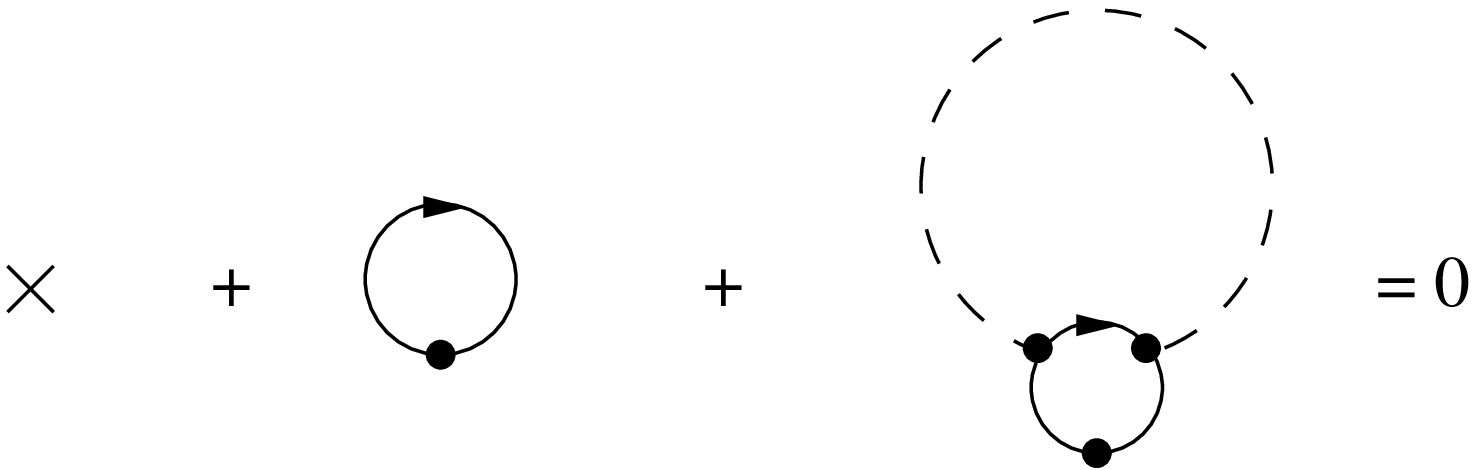}}
\caption{The diagrams contributing to the stationary-point equation
 ({{\protect{\ref
{ge2}}}}) with meson loops included. The dashed line corresponds to the
meson propagators $K_{\pi }$ and $K_{\sigma }$ defined in Eq.~({{\protect{\ref
{eq:pi3}}}}). The cross denotes the term~$a^{2}S_{0}$.}
\label{fig:GE1}
\end{figure}
In particular, meson loops are finite, and we now incorporate them in our
study.
First we analyze the problem at $T=0$. The meson-loop contribution in the
effective action can be obtained in a standard way by integrating out meson
fluctuations around the stationary-point configuration \cite{loops:NJL}. The
new stationary-point condition for $S_{0}$ assumes the form 
\begin{eqnarray}
&&S_{0}\left( a^{2}-2\nu g(S_{0})-2\nu \frac{m}{S_{0}}g^{\prime
}(S_{0})\right)  \nonumber \\
&&~~~~~~~~+{\frac{1}{2}}\int \frac{d^{4}p}{(2\pi )^{4}}\,\left( V_{\sigma
}(p)K_{\sigma }(p)+N_{B}V_{\pi }(p)K_{\pi }(p)\right) =0\;,  \label{ge2}
\end{eqnarray}
where $N_{B}=N_{f}^{2}-1=3$ is the number of pions, $K_{\sigma }$ and $%
K_{\pi }$ are meson propagators at the quark-loop level and the 3-point
quark loops (see Fig. 2) are equal to: 
\begin{eqnarray}
V_{\sigma }(p) &=&-2\nu \int \frac{d^{4}k}{(2\pi )^{4}}\frac{R^{2}(p)R(k+p)}{%
D^{2}(k)D(k+p)}(-2{\cal M}(k)\,k\cdot (k+p)  \nonumber \\
&-&{\cal M}(k+p)(k^{2}+{{\cal M}^{2}(k)})) \\
V_{\pi }(p) &=&-2\nu \int \frac{d^{4}k}{(2\pi )^{4}}\frac{R^{2}(k)R(k+p)}{%
D^{2}(k)D(k+p)}(-2{\cal M}(k)\,k\cdot (k+p)  \nonumber \\
&+&{\cal M}(k+p)(k^{2}-{{\cal M}^{2}(k)}))  \label{tr1box}
\end{eqnarray}
The inverse meson propagators can be written as: 
\begin{eqnarray}
K_{\pi }^{-1}(p) &=&\nu p^{2}f_{\pi }(p)+\frac{m}{S_{0}}2\nu g^{\prime
}(S_{0})\;,  \nonumber \\
K_{\sigma }^{-1}(p) &=&K_{\pi }^{-1}(p)+4\nu S_{0}^{2}f_{\sigma }(p)\;.
\label{eq:pi3}
\end{eqnarray}
where we have introduced 
\begin{eqnarray}
f_{\pi }(p) &=&-\frac{2}{p^{2}}\int \frac{d^{4}k}{(2\pi )^{4}}\left( \frac{%
R(k)R(k+p)(k\cdot (k+p)+{\cal M}(k){\cal M}(k+p))}{D(k)D(k+p)}\right. 
\nonumber \\
&&~~~~~~~~~~~~~\qquad ~~\left. -\frac{R^{2}(k)}{D(k)}\right) \;
\end{eqnarray}
and 
\begin{equation}
f_{\sigma }(p)=\frac{1}{S_{0}^{2}}\int \frac{d^{4}k}{(2\pi )^{4}}\frac{%
R(k)R(k+p){\cal M}(k){\cal M}(k+p)}{D(k)D(k+p)}\;.  \label{fs}
\end{equation}
All terms included in the stationary-point condition are shown in Fig. 2. A
similar expression can be obtained for the quark condensate.

The results are somewhat surprising. We have found large cancellations in
the meson-loop contribution, such that the net correction to the value of $%
S_{0}$ is less than 1\%, and to $\langle \bar{q}q\rangle $ of the order of
3-4\% on top of the quark-loop value. To be more precise, for the Gaussian
regulator with the parameters $\Lambda =754{\rm MeV}$, $a=176{\rm MeV}$, and 
$m=0$ we find $\langle \bar{q}q\rangle =(0.0505+0.0021-0.0041){\rm GeV}^{3}$%
, where we have decomposed in the quark-loop, $\sigma $-loop, and the
pion-loop contribution, respectively. We note the cancellation between the $%
\sigma $- and pion-loop contributions. The phenomenon occurs for various
regulators and parameter sets. It is also instructive to look at the
integrand of the $p$-integration in Eq. (\ref{ge2}). After carrying the
angular integration, we are left with a radial integration over $|\vec{p}|$.
The integrand in the pion-loop case changes sign as we increase $|\vec{p}|$,
and this results in an additional cancellation. At large values of $|\vec{p}|
$ the integrand drops to zero. This behavior is quite different from the
calculations in the local model, as in Ref. \cite{loops:NJL}. There the
analogous integrands in the meson-loop contributions increase with $|\vec{p}|
$, and the results depend strongly on the value of the momentum cut-off in
the meson loop. 

Having found small meson-loop effects for $T=0$ we expect them to remain
small at finite $T$. This is indeed the case. The expression (\ref{ge2}) and
the similar expression for the quark condensate can be thermalized by the
prescription (\ref{temp}). The only difference now is that in the mesonic
loop we have to use bosonic Matsubara frequencies,{\em \ i.e. }in the
meson-loop we replace 
\begin{equation}
\int \frac{d^{4}p}{\left( 2\pi \right) ^{4}}f(p)=\int \frac{dp_{0}}{2\pi }%
\int \frac{d^{3}p}{\left( 2\pi \right) ^{3}}f(p_{0},\vec{p})\longrightarrow
T\sum_{j=-\infty }^{\infty }\int \frac{d^{3}p}{\left( 2\pi \right) ^{3}}%
f(\omega _{j},\vec{p}),  \label{tempbos}
\end{equation}
with $\omega _{j}=2\pi jT$. The numerical calculations show that the effect
of the meson loops for finite temperatures is tiny, and the quark loop is
dominant. The critical temperatures are practically unchanged. This is
different than in local models, where the meson loops have a substantial
effect on the finite-$T$ behavior of the model \cite{Temp}.

Our main conclusions are as follows: the restoration of chiral symmetry in
non-local chiral quark models proceeds similarly to the conventional
Nambu--Jona-Lasinio models. In the strict chiral limit the transition is
second order, however the critical temperatures are lower, around $100{\rm %
MeV}$. For finite current quark masses we find a smooth cross-over to the
restored phase. An interesting result is a very small contribution of meson
loops to the quark condensate, which is in contrast to the calculations
within the Nambu--Jona-Lasinio model. The effect occurs for a variety of
regulators used. It is encouraging, since it indicates that we can rely on
the simple-to-do one-quark-loop calculations. It remains to be seen whether
in non-local models the meson loop effects are also negligible for meson
correlators and other Green's functions. 

\ack
One of us (WB) is grateful to Georges Ripka for many discussions on the
topics of this paper.


\end{document}